# REAK: Reliability analysis through Error rate-based Adaptive Kriging

Zeyu Wang and Abdollah Shafieezadeh[1]
The Ohio State University, Columbus, OH, 43202, United States

**ABSTRACT**
As models in various fields are becoming more complex, associated computational demands have been increasing significantly. Reliability analysis for these systems when failure probabilities are small is significantly challenging, requiring a large number of costly simulations. To address this challenge, this paper introduces Reliability analysis through Error rate-based Adaptive Kriging (REAK). An extension of the Central Limit Theorem based on Lindeberg condition is adopted here to derive the distribution of the number of design samples with wrong sign estimate and subsequently determine the maximum error rate for failure probability estimates. This error rate enables optimal establishment of effective sampling regions at each stage of an adaptive scheme for strategic generation of design samples. Moreover, it facilitates setting a target accuracy for failure probability estimation, which is used as stopping criterion for reliability analysis. These capabilities together can significantly reduce the number of calls to sophisticated, computationally demanding models. The application of REAK for four examples with varying extent of nonlinearity and dimension is presented. Results indicate that REAK is able to reduce the computational demand by as high as 50% compared to state-of-the-art methods of Adaptive Kriging with Monte Carlo Simulation (AK-MCS) and Improved Sequential Kriging Reliability Analysis (ISKRA).

## 1. INTRODUCTION

Engineered structures and systems as well as those found in nature are vulnerable against various forms of stressors that threaten their functionality and integrity. Reliability analysis enables analyzing the performance of structures and systems in terms of the probability of them failing to meet a prescribed objective considering aleatoric and epistemic uncertainties in structures or systems and their stressors. Generally, the analysis of reliability is conducted based on a limit state function $g(X)$, the response of which determines the state of the system: $g(X) \leq 0$ indicates failure and $g(X) > 0$ means survival; the boundary region where $g(X) = 0$ is called the limit state. Thus, the probability of failure can be defined as:

$$P_f = P(g(X) \leq 0) \qquad (1)$$

Several methodologies have been developed in order to calculate $P_f$. Crude Monte Carlo simulation (MCS) [1], [2], First and Second Order Reliability Methods (FORM & SORM) [3], [4], Importance Sampling [5] and Subset Simulation [6] are among such techniques. Based on the Monte Carlo simulation approach, which is often regarded as the benchmark to compare the accuracy and efficiency of other methods, the probability of failure is calculated as follows, as long as the number of design points, $N_{MCS}$, is sufficiently large:

$$P_f \cong P_f^{MCS} = \frac{N_{g(X) \leq 0}}{N_{MCS}} \qquad (2)$$

where $N_{g(X) \leq 0}$ is the number of design points that satisfy $g(X) \leq 0$ out of $N_{MCS}$ total design points. Despite the high accuracy that can be achieved using MCS method, the required significantly large number of calls to the limit state function presents a major challenge for many applications where evaluation of structure or system performance is very costly. Therefore, it is imperative to devise strategies and develop techniques that can reduce the number of calls to limit state functions while maintaining desired levels of accuracy in the estimation of $P_f$. FORM and SORM [7], by linearizing the performance function in the vicinity of the most probable failure point (MPP), present approximate solutions for reliability analysis problems [7]. However, these methods fail to provide accurate estimates of $P_f$ in problems with highly nonlinear limit state functions. Importance Sampling [5], [8], Subset Simulation [6], [9] and other similar techniques improve the efficiency of reliability estimation compared to crude MCS method; however due to lack of strategic selection of design points, referred to as points where limit state function is evaluated, they are not among the most efficient techniques.

---

[1] Corresponding author: Shafieezadeh.1@osu.edu

In recent years, meta-models have been used to enhance reliability analysis procedures based on the premise that a surrogate model generated using a limited number of design points can replace the original complex and time consuming model [10], [11] in reliability estimation procedures. Several meta-modeling techniques including Polynomial Response Surface [12]–[14], Polynomial Chaos Expansion (PCE) [15], Support Vector Regression (SVR) [10], [16], and Kriging [7] have been used for this purpose. The latter approach is the most popular as it provides estimations of expected responses and their variance over the space of random variables in the limit state function by assuming that the true responses of the model and the estimated ones follow a joint Gaussian distribution. The stochastic characterization feature of Kriging [17], [18] facilitates quantification of the confidence around estimated responses, which enables active learning. Efficient Global Optimization (EGO) [19] was among the first algorithms that leveraged this feature of Kriging to identify valuable points for generation of surrogate Kriging models. However, EGO does not account for the importance of regions in sample space for the reliability estimation; for example, it does not prioritize samples where the response of the performance function is close to the limit state $g(X) = 0$, despite the fact that these points are expected to considerably contribute to the failure probability. To overcome this shortcoming, Efficient Global Reliability Analysis (EGRA) was proposed [20] which offers a suitable learning function for reliability analysis problems. This learning function, called Expected Feasibility Function ($EFF$), prioritizes points among candidates that have large variance, and are in the vicinity of the limit state. The reliability estimation process via EGRA can be summarized as follows: first, generate candidate design samples and initial training samples using Latin Hypercube sampling ($LHS$) in probability space ($\pm$ five standard deviations) with prescribed number of $\frac{(n_{ran}+1)(n_{ran}+2)}{2}$ for initial training samples, in which $n_{ran}$ is the number of random variables. Second, evaluate the limit state function for these initial training samples and establish the initial Kriging model. Then, search among candidate design samples for the point that maximizes $EFF$ in the candidate design space, add this sample to the set of training samples and evaluate the performance function. Next, construct the refined Kriging model and repeat previous steps until convergence to the stopping criterion (usually $max(EFF) < 0.001$). Finally, use the resulting Kriging model to estimate the probability of failure through Monte-Carlo simulation. One of the main drawbacks of EGRA is the lack of a logical and systematic approach to define the number of candidate design samples, especially for high-dimensional problems. Moreover, the estimated probability of failure may not be sufficiently accurate due to the possibility that the number of candidate design samples for MCS may not be large enough [7]. To address the latter limitation, Echard et al. [7] introduced an adaptive algorithm called Adaptive Kriging with Monte Carlo Simulation (AK-MCS) that recursively increases the number of candidate design samples for MCS until the convergence criterion defined based on the covariance of $P_f$ is met. Two learning functions, $EFF$ and $U$, have been used in AK-MCS; both functions are found to produce acceptable results [7], [21]. However, $EFF$ convergences faster to the true probability of failure compared to $U$, and $U$ convergences faster to its stopping criterion ($min(U) \geq 2$) than $EFF$ ($max(EFF) < 0.001$). To improve AK-MCS, a learning function called Least Improvement Function ($LIF$) was proposed in [22] to take the probability density function (pdf) of each candidate design sample into consideration. Wen et al. [23] points out that AK-MCS, fails to completely eliminate the candidate design samples with weak probability density, if the number of candidate design samples is large, and proposed the Sequential Kriging reliability analysis (ISKRA) method. The approach is based on the premise that candidate design points with low probability density have negligible impact on the accuracy of failure probability estimations; therefore, they can by entirely neglected in reliability analysis.

Despite improvements offered in previous studies, a major limitation still persists concerning the stopping criterion; the stopping criteria, including but not limited to $max(EFF) < 0.001$ and $min(U) \geq 2$ are set arbitrarily without direct connection to the accuracy of reliability estimates. In the lack of such links and the ability to quantify error in failure probability estimates, and to ensure that the accuracy of produced results is acceptable, former investigations chose very strict thresholds for $EFF$ and $U$ metrics, among others. This can potentially result in setting extremely high accuracy levels for the error rate in failure probability for many cases. There always exists the potential of 'overfitting' for Kriging models, where the true error rate estimated by MCS, $\epsilon$, is close to zero; however the algorithms based on $EFF$ and $U$ stopping criteria continue renewing the model, thus unnecessarily increasing the number of calls to limit state functions. Moreover, in many engineering applications, looser error thresholds may be acceptable in order to strike a balance between accuracy and computational costs; however, the current stopping criteria fail to address this need.

This paper introduces an analytical approach to derive an approximate upper bound for failure probability estimates in Kriging-based simulation methods; the upper bound here is referred to as the maximum error rate $\hat{\epsilon}_{max}$. Inaccuracies in estimating the sign (+ or -) of the performance function for candidate design samples in MCS process substantially contributes to the overall error rate. By applying the Central Limit Theorem [24] and its extension for non-identical random variables based on Lindeberg condition [25], the distribution of the number of design samples with wrong sign estimations is mathematically derived and used to determine $\hat{\epsilon}_{max}$. This new capability is then implemented as a stopping criterion in the form of $\epsilon \leq \hat{\epsilon}_{max} \leq \epsilon_{thr}$ to replace conventional criteria. This offers the flexibility to set the threshold of error rate, $\epsilon_{thr}$, according to the requirement of the problem at hand. Generally, the number of calls to the performance function, denoted as $N_{call}$, is inversely proportional to $\epsilon_{thr}$. Reliability analysis through Error rate-based Adaptive Kriging (REAK) method is then proposed in which candidate design samples are added sequentially according to their probability density and using the estimated maximum error rate $\hat{\epsilon}_{max}$ as the stopping criterion. Another major enhancement offered by REAK is that, in the learning process, REAK initially neglects the candidate deign points that their probability density is under threshold value $\rho_{thr}$. This idea, first proposed by Wen et al. [23], has been shown to considerably reduce $N_{call}$; however, the definition of $\rho_{thr}$ in [23] is arbitrary and remains fixed throughout the process of failure probability estimation. If more candidate design points are neglected, $N_{call}$ will be reduced at the expense of a slightly higher error rate, meaning that $\epsilon \propto \rho_{thr}$ & $\epsilon \propto \frac{1}{N_{call}}$. Accordingly, REAK initially sets $\rho_{thr}$ to be large to only take those 'important' candidate design samples into consideration, then gradually reduces $\rho_{thr}$ until the stopping criterion is met. By adaptively expanding the effective sampling regions, REAK effectively neglects the candidate design samples with weak probability density in an optimal manner, which can considerably reduce the number of calls to the performance function and therefore substantially reduce computational demands.

This paper is organized in four sections. The Kriging model and the corresponding AK-MCS algorithm are briefly introduced in Section 2. In Section 3, the definition of effective sampling region, the mathematical derivation of estimated maximum error rate $\hat{\epsilon}_{max}$ and the proposed REAK method are presented. In Section 4, the application of REAK for four examples with varying extent of nonlinearity and dimensions is presented and the results are compared with those from MCS, AK-MCS and ISKRA methods. Section 5 presents the conclusions of this study.

## 2. AK-MCS APPROACH

To overcome the shortcoming of significantly large computational times in simulation-based reliability analysis techniques, AK-MCS uses a learning function ($EFF$ or $U$) to strategically select valuable points to adaptively refine the embedded Kriging model [7]. The approach also uses iteration stopping criteria to determine whether produced estimates of reliability analysis have acceptable accuracy. The following subsections provide a brief review of Kriging, learning functions, and reliability estimation using AK-MCS.

### 2.1 Kriging Model

The Kriging model, also called Gaussian Process Regression, makes a prior assumption that the estimated response $\hat{y}(x)$ and the known true response $y$ follow a joint Gaussian distribution [26], [27]. Based on this assumption, Kriging combines the process of interpolation and regression. The estimated stochastic response $K(x)$ for input $x$ can be described as follows:

$$K(x) = \beta^T f(x) + Z(x, w) \quad (3)$$

where $f(x)$ is the basis function and $\beta$ is the vector of regression coefficient of $f(x)$. $\beta^T f(x)$ that represents the mean value of $K(x)$, which is often assumed to have ordinary ($\beta_0$), linear ($\beta_0+\sum_{i=1}^{N}\beta_i x_i$) or quadratic ($\beta_0+\sum_{i=1}^{N}\beta_i x_i+\beta_0+\sum_{i=1}^{N}\sum_{j=1}^{N}\beta_{ij}x_i x_j$) forms, where $N$ is the dimension of the random input vector $x$. More details on $f(x)$ and $\beta$ in Kriging models can be found in [27]. In this study, the ordinary Kriging model is used, meaning that both $f(x)$ and $\beta$ are constant. $Z(x)$ is a stationary normal Gaussian process with zero mean and the following covariance matrix:

$$COV(Z(x), Z(w)) = \sigma^2 R(x, w, \theta) \quad (4)$$

where $x$ and $w$ are two arbitrary points, and $\sigma^2$ is the process variance, and represents the generalized mean square error in the regression process. Moreover, $R(x, w, \theta)$, called kernel function, represents the correlation function of the process with hyper-parameter $\theta$. A set of correlation functions have been implemented in Kriging including, but not limited to, Linear, Exponential, Gaussian and Matérn functions. In this article, the separable anisotropic Gaussian function is used which has the following form:

$$R(x, w, \theta) = \prod_{i=1}^{N} \exp(-\theta_i (x_i - w_i)^2) \quad (5)$$

The hyper-parameter $\theta$ can be determined using methods such as Maximum Likelihood Estimation (MLE) and Cross-Validation (CV) [27], among others. In order to be consistent with previous studies for comparison purposes, here, $\theta_i$ is searched in (0,10) using MATLAB optimization toolbox DACE [28], [29] that uses MLE method. The Maximum Likelihood Estimation approach is described below:

$$\theta = \underset{\theta^* \in \Theta}{\mathrm{argmin}} \left( |R(x, w, \theta)|^{\frac{1}{m}} \sigma^2 \right) \quad (6)$$

In the Kriging model, the regression coefficient $\beta$, estimated mean response and variance can be determined as follows:

$$\beta = (F^T R^{-1} F)^{-1} F^T R^{-1} y \quad (7)$$

$$\mu_K(x) = f^T(x)\beta + r^T(x)R^{-1}(y - F\beta) \quad (8)$$

$$\sigma_K^2(x) = \sigma^2(1 - r^T(x)R^{-1}r(x) + (F^T R^{-1}r(x) - f(x))^T (F^T R^{-1} F)^{-1}(F^T R^{-1}r(x) - f(x))) \quad (9)$$

where $F$ is the matrix of basis functions $f(x)$ evaluated at known training points, i.e. $F_{ij} = f_j(x_i)$, $i = 1, 2, ..., m; j = 1, 2, ..., p$. $r(x)$ is the vector of correlations between known training points $x_i$ and an unknown point $x$: $r_i = R(x, x_i, \theta)$, $i = 1,2 ... m$. $R$ is the autocorrelation matrix for known training points: $R_{ij} = R(x_i, x_j, \theta)$, $i = 1,2, ..., m; j = 1,2, ..., m$. The stochastic response $K(x)$ can then be represented using a normal distribution as:

$$K(x) \sim N\left(\mu_K(x), \sigma_K^2(x)\right) \quad (10)$$

According to this model, response predictions of points close to known training points will have higher confidence compared to those that are further away from the training points. The probabilistic information provided by the Kriging model including the expected value of predictions and their variance can be leveraged to select next evaluation points in the reliability estimation more effectively. This Kriging statistical property has been used in AK-MCS for sequential selection of training points for model refinement as explained further below.

## 2.2. Learning Function

Learning functions have a crucial role in AK-MCS. As the name implies, the 'learning' refers to the process of iterative selection of points for Kriging refinement based on the stochastic information for each design point. Two popular forms of learning function are the Expected Learning Function ($EFF$) and $U$ function. $EFF$ in Efficient Global Reliability Analysis (EGRA) algorithm [20], [30], tends to select points that are close to the limit state $g(X) = a$ and/or have high variance. The mathematical expression of $EFF$ is presented below:

$$\begin{aligned}EFF(x) = (\mu_K(x) - a)&\left[2\Phi\left(\frac{a - \mu_K(x)}{\sigma_K(x)}\right) - \Phi\left(\frac{a^- - \mu_K(x)}{\sigma_K(x)}\right) - \Phi\left(\frac{a^+ - \mu_K(x)}{\sigma_K(x)}\right)\right]\\ -\sigma_K(x)&\left[2\phi\left(\frac{a - \mu_K(x)}{\sigma_K(x)}\right) - \phi\left(\frac{a^- - \mu_K(x)}{\sigma_K(x)}\right) - \phi\left(\frac{a^+ - \mu_K(x)}{\sigma_K(x)}\right)\right]\\ +2\sigma_K(x)&\left[\Phi\left(\frac{a^+ - \mu_K(x)}{\sigma_K(x)}\right) - \Phi\left(\frac{a^- - \mu_K(x)}{\sigma_K(x)}\right)\right]\end{aligned} \quad (11)$$

where $\phi(\cdot)$ is the standard normal probability density function (PDF) and $\Phi(\cdot)$ is the standard normal cumulative density function (CDF). In this paper, $a = 0$, $a^+ = 2\sigma_K(x)$ and $a^- = -2\sigma_K(x)$. The point that maximizes the $EFF$ response is chosen as the next point to refine Kriging model. The iteration stopping criterion is usually expressed as $max(EFF(x)) \leq 0.001$. It is always true that the accuracy of the estimation of the probability of failure increases as the threshold of $EFF$ stopping criterion decreases at the expense of increasing the computational demand.

Another powerful learning function is $U$ that is concerned with uncertainties in sign ($\pm$) estimation of $\hat{g}(x)=0$. In this regard, $U$ takes the probabilistic distribution of estimated responses into consideration, and quantifies the probability of making a wrong sign estimation in $\hat{g}(x)$. The formulation of $U$ is:

$$U(x) = \frac{|\mu_K(x)|}{\sigma_K(x)} \quad (12)$$

The probability of making a wrong sign estimation is:

$$P_w(x) = 1 - \Phi(U(x)) = \Phi\left(-\frac{|\mu_K(x)|}{\sigma_K(x)}\right) \quad (13)$$

The point that minimizes the response of $U$ is selected in the learning iteration. The stopping criterion for $U$ learning function is often set as $min(U(x)) \geq 2$, which is interpreted as the probability of making wrong sign estimation to be less than 0.023. It is shown that both the $EFF$ and $U$ learning functions are efficient in selection of design points [7]. However, $EFF$ tends to converge faster than $U$ in achieving true probability of failure $P_f$, while $U$ learning function converges faster to its own stopping criterion $(min(U(x)) \geq 2)$ than $EFF$ $(max(EFF(x)) \leq 0.001)$. In this paper, learning function $EFF$ is used.

## 2.3. Reliability Estimation Using AK-MCS

The general principle of AK-MCS is to start with a small number of candidate design samples to predict $\hat{P}_f$ and then adaptively increase the number of candidate design samples $N_{MCS}$ until the stopping criterion is satisfied. The stopping criterion for this process in AK-MCS is:

$$COV_{P_f} = \sqrt{\frac{1 - \hat{P}_f}{\hat{P}_f N_{MCS}}} \leq COV_{\text{thr}} \quad (14)$$

where $\hat{P}_f$ is the estimated failure probability by AK-MCS, $COV_{P_f}$ is the coefficient of variation of $\hat{P}_f$, $N_{MCS}$ is the number of candidate design points for Monte Carlo simulation and $COV_{\text{thr}}$ is the threshold for AK-MCS stopping criterion. It is obvious that $N_{MCS}$ considerably influences $COV_{P_f}$, which means the final $\hat{P}_f$ is reliable only when $N_{MCS}$ is large enough so that $COV_{P_f} \leq COV_{\text{thr}}$. The flowchart of AK-MCS is presented in Fig. 1.

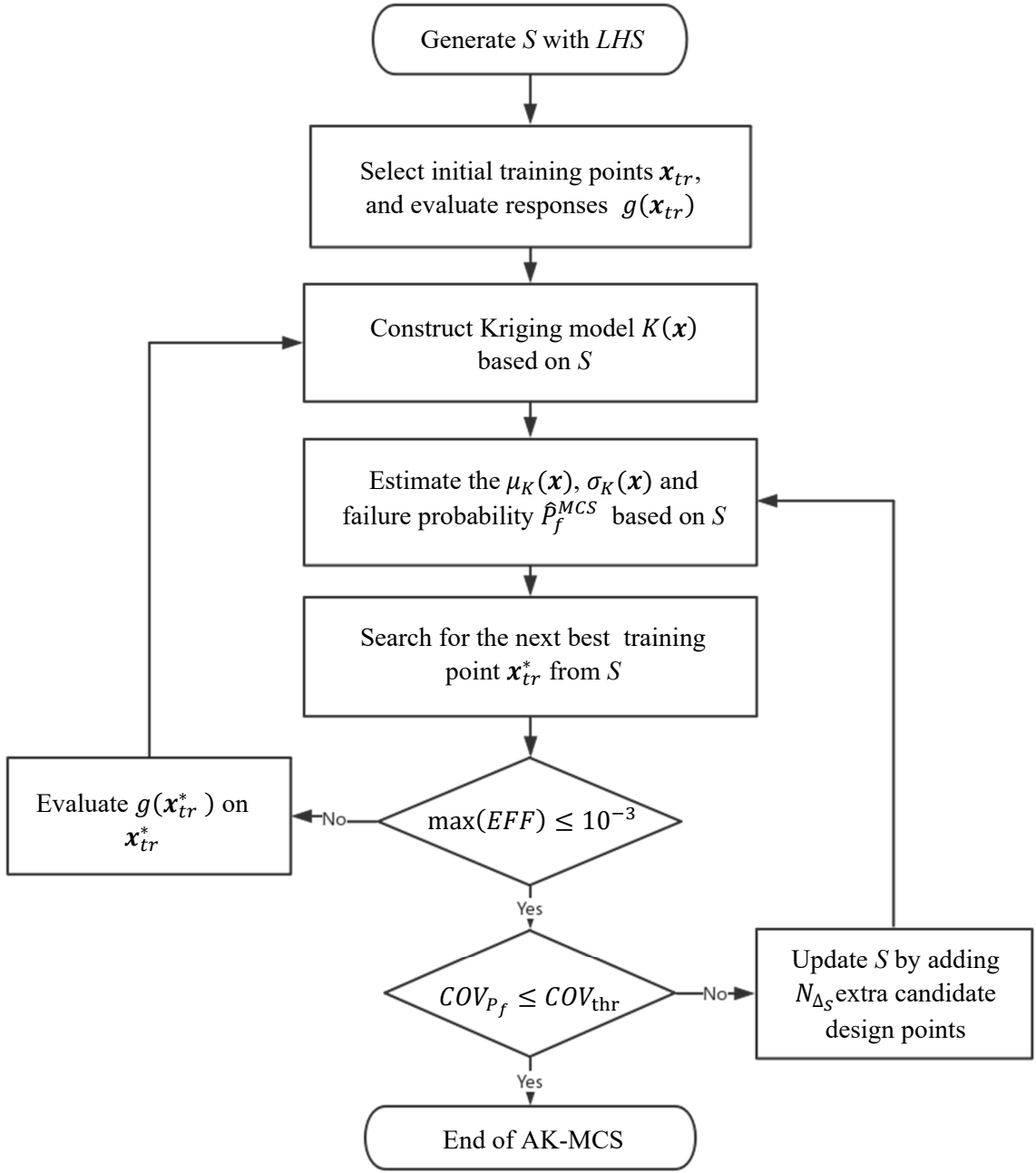

**Fig** 1. AK-MCS flowchart

## 3. REAK: THE PROPOSED METHOD
To predict the probability of failure, meta-model based reliability analysis algorithms employ sampling techniques such as crude MCS, importance sampling and subset simulation [5]–[7], [10], [23]. The relative error of the estimated probability of failure with respect to that derived from MCS can be defined as:

$$\epsilon = \frac{|\hat{P}_f - P_f^{MCS}|}{P_f^{MCS}} = \left|\frac{\hat{P}_f}{P_f^{MCS}} - 1\right| \tag{15}$$

where $\hat{P}_f$ is the failure probability estimated by the meta-model and $P_f^{MCS}$ is the one estimated by crude Monte-Carlo simulation. In order for the error rate to make sense, the set of candidate design samples for crude MCS and Kriging surrogate model based MCS should be the same. Thus, a reliability analysis algorithm $\Psi$ is optimal, if it requires the least number of calls to the performance function, $N_{call}$, while maintaining the error rate smaller than a prescribed threshold $\epsilon_{thr} \in (0,1)$:

$$\Psi^* = \underset{\Psi \in \Gamma, \ \epsilon \leq \epsilon_{thr}}{\arg\min} \ N_{call} \tag{16}$$

where $\Gamma$ is the set of all feasible algorithms. In this context, this paper proposes Reliability analysis through Error rate-based Adaptive Kriging (REAK) for reliability assessment. REAK integrates two novel features including effective sampling regions (ESR) and maximum error rate prediction with AK-MCS.

### 3.1. Effective Sampling Regions
This section presents effective sampling regions which is developed by sequentially updating the adaptive sampling regions (ASR) [23]. Adaptive sampling regions denote those random realization points that have probability density larger than a threshold value $\rho_{thr}$. Neglecting the points outside the domain specified by ASR, Wen et al. [23] reduced $N_{call}$ because candidate design samples with small probability density have negligible contribution to $\hat{P}_f$. $\rho_{thr}$ in this approach is established as:

$$P_{\{\rho(x) > \rho_{thr}\}} = \alpha \hat{P}_f^{n-1} \tag{17}$$

where $\rho(x)$ is the joint probability density of candidate design samples $x$, $\alpha$ is a constant coefficient and $\hat{P}_f^{n-1}$ is the probability of failure achieved by Kriging model of ISKRA [23] in the last iteration. The regions satisfying the equation above are called adaptive sampling regions. Importantly, the coefficient $\alpha$ affects the accuracy of $\hat{P}_f$ as well as $\epsilon$. It is stated that the error rate $\epsilon$ (defined in Eq. 15) is always smaller than $\alpha$ even in the worst scenario, if the surrogate model in ASR is refined enough [23]. It should be noted that $N_{call}$ decreases considerably as $\alpha$ increases.

The selection of $\alpha$ in [23] is empirical. Although setting $\alpha \in [0.05, 0.2]$ often results in acceptable reliability estimates [23], the selection of $\alpha$ in this manner may not be optimal. For instance, it is shown that even $\alpha > 1$ may result in error rates smaller than 0.05, which can result in significant reduction in $N_{call}$. This motivates developing an approach that logically and quantitatively search for the optimal $\alpha$ so that the true error rate $\epsilon$ is close to but slightly smaller than the threshold error rate $\epsilon_{thr}$. For this purpose, this paper proposes effective sampling regions that are achieved by sequentially decreasing $\alpha$ from an initially high value to the optimal one. The illustration of ESR is presented in Fig 2. Through implementation of ESR, the next optimal training point $x_{tr}^*$ can be determined by maximizing $EFF$ as follows:

$$x_{tr}^* = \underset{x \in ESR}{\arg\max} \ EFF(x) \tag{18}$$

For learning purposes, the above optimization problem can be converted to one without constraint $x \in ESR$ as follows [23]:

$$EFF_m(x) = \begin{cases} EFF(x), & x \in ESR \\ 0, & otherwise \end{cases} \tag{19}$$

where $EFF_m$ stands for the modified $EFF$ learning function. The methodology of determining optimal $\alpha$ is presented in section 3.3 after the theoretical derivation of estimated maximum error rate $\hat{\epsilon}_{max}$ ($\epsilon \leq \hat{\epsilon}_{max} \leq$

$\epsilon_{thr}$) in section 3.2.

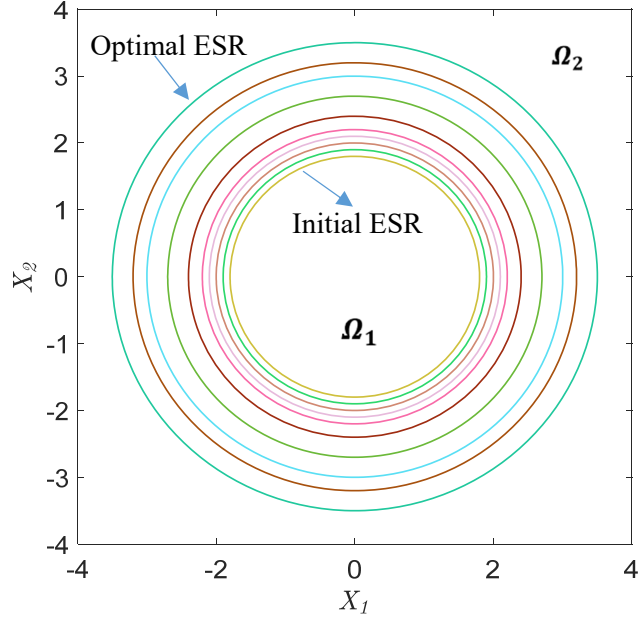

**Fig** 2. Effective sampling regions for the probability space formed by two independent normal random variables.

### 3.2. Maximum Error Rate Estimation

In this subsection, the approximate upper bound or maximum true error rate $\hat{\epsilon}_{max}$ is derived using the Central Limit Theorem for non-identical random variables based on Lindeberg condition [25]. Given $\alpha$, the entire sample space $\Omega$ can be divided into mutually exclusive and collectively exhaustive sets $\Omega_1$ and $\Omega_2$ (shown in Fig. 2) that refer to the space of design samples that satisfy $P_{\{\rho(x)>\rho_{thr}\}} = \alpha \hat{P}_f^{n-1}$ and $P_{\{\rho(x)\leq\rho_{thr}\}} = \alpha \hat{P}_f^{n-1}$, respectively. Let's define $\Omega_1^f$ and $\Omega_1^{\hat{f}}$ as subsets of $\Omega_1$ with failed design samples according to $g(\cdot)$ and $K(\cdot)$, respectively, and $\Omega_1^s$ and $\Omega_1^{\hat{s}}$ as complements of $\Omega_1^f$ and $\Omega_1^{\hat{f}}$, respectively in $\Omega_1$. $\Omega_2^f, \Omega_2^{\hat{f}}, \Omega_2^s$ and $\Omega_2^{\hat{s}}$ are defined similarly but over $\Omega_2$. Moreover, $N$ is used here to indicate the size of a set. For example, $N_{\Omega_1^f}$ indicates the size of $\Omega_1^f$ or, in another word, the number of points in $\Omega_1$ that are evaluated as failed points according to $g(\cdot)$. Based on these notations, the estimated probability of failure can be expressed as:

$$\hat{P}_f = \frac{N_{\Omega_1^{\hat{f}}} + N_{\Omega_2^{\hat{f}}}}{N_\Omega} \tag{20}$$

$$P_f^{MCS} = \frac{N_{\Omega_1^f} + N_{\Omega_2^f}}{N_\Omega} \tag{21}$$

The error rate can be calculated as:

$$\epsilon = \left|\frac{\hat{P}_f}{P_f^{MCS}} - 1\right| = \left|\frac{N_{\Omega_1^{\hat{f}}} + N_{\Omega_2^{\hat{f}}}}{N_{\Omega_1^f} + N_{\Omega_2^f}} - 1\right| \tag{22}$$

Rearranging the above equation yields:

$$\epsilon = \left| \frac{N_{\Omega_1^{\hat{f}}} + N_{\Omega_2^{\hat{f}}} + N_{\Omega_2^{\hat{f}}} - N_{\Omega_2^{f}}}{N_{\Omega_1^{f}} + N_{\Omega_2^{f}}} - 1 \right| = \left| \frac{N_{\Omega_2^{\hat{f}}} - N_{\Omega_2^{f}}}{N_{\Omega_1^{f}} + N_{\Omega_2^{f}}} + \frac{N_{\Omega_1^{\hat{f}}} + N_{\Omega_2^{f}}}{N_{\Omega_1^{f}} + N_{\Omega_2^{f}}} - 1 \right| \tag{23}$$

When the threshold of the stopping criterion for learning function $EFF$ is set to be small, $N_{\Omega_1^{\hat{f}}}$ will be very accurate. Given that this stopping criterion is often expressed as $max(EFF(x)) \leq 0.001$, it can be concluded that:

$$\lim_{max(EFF(x)) \to 0.001} N_{\Omega_1^{\hat{f}}} \cong N_{\Omega_1^{f}}, \quad x \in \Omega_1 \tag{24}$$

If the stopping criterion is set in every literation, then:

$$N_{\Omega_1^{\hat{f}}} \cong N_{\Omega_1^{f}} \tag{25}$$

Then combining Eq. (23) and Eq. (25) results in:

$$\epsilon \cong \left| \frac{N_{\Omega_2^{\hat{f}}} - N_{\Omega_2^{f}}}{N_{\Omega_1^{\hat{f}}} + N_{\Omega_2^{f}}} \right| \tag{26}$$

For each design point $x_i \in \Omega_2^{\hat{s}}$, let's define an indicator function $I_i^{\hat{s}}$ that takes one when $K(x_i)$ makes a wrong estimation of the sign of $g(x_i)$, and zero when the sign estimation is correct. The probability of the event that the sign estimate of $x_i$ is wrong can be determined as:

$$P(I_i^{\hat{s}} = 1 \mid x_i \in \Omega_2^{\hat{s}}) = P_i = \Phi\left(-\left|\frac{\hat{y}_{x_i}}{\hat{\sigma}_{x_i}}\right|\right) \tag{27}$$

$I_i^{\hat{s}}$ in fact follows a Bernoulli distribution with the following mean and variance:

$$E[I_i^{\hat{s}} \mid x_i \in \Omega_2^{\hat{s}}] = P_i, \quad Var[I_i^{\hat{s}} \mid x_i \in \Omega_2^{\hat{s}}] = P_i(1 - P_i) \tag{28}$$

Thus, the total number of $x_i$ with wrong sign estimate can be defined as $N_{\Omega_2^{\hat{s}}}^{WSE} = \sum_{i=1}^{N_{\Omega_2^{\hat{s}}}} I_i^{\hat{s}}$. Considering the statistical independence of $x_i$,

$$\mu_{\Omega_2^{\hat{s}}} = E\left[N_{\Omega_2^{\hat{s}}}^{WSE}\right] = \sum_{i=1}^{N_{\Omega_2^{\hat{s}}}} P_i, \quad \sigma_{\Omega_2^{\hat{s}}}^2 = Var\left[N_{\Omega_2^{\hat{s}}}^{WSE}\right] = \sum_{i=1}^{N_{\Omega_2^{\hat{s}}}} P_i(1 - P_i) \tag{29}$$

Given that

$$\lim_{N_{\Omega_2^{\hat{s}}} \to \infty} \left( \max_{i=1,\dots,N_{\Omega_2^{\hat{s}}}} \frac{Var[I_i^{\hat{s}}]}{Var\left[N_{\Omega_2^{\hat{s}}}^{WSE}\right]} \right) = 0 \tag{30}$$

Lindeberg's condition for the Central Limit Theorem for non-identical and independent random variables is satisfied. It should be noted that events of wrong sign estimation of candidate design points by Kriging model are correlated, and considering these correlations is expected to enhance the accuracy of the maximum error rate. However, including correlations for candidate design points in the framework significantly increases the computational complexity and demands. Due to this and other reasons explained later in the paper, this study considers independent responses. It then follows that when $N_{\Omega_2^{\hat{s}}}$ is sufficiently large, $N_{\Omega_2^{\hat{s}}}^{WSE}$ in distribution converges to a normal distribution:

$$N_{\Omega_2^{\hat{s}}}^{WSE} \sim N\left(\mu_{\Omega_2^{\hat{s}}}, \sigma_{\Omega_2^{\hat{s}}}^2\right) \tag{31}$$

However, the total number of design samples in $\Omega_2^{\hat{f}}$ with wrong sign estimate, $N_{\Omega_2^{\hat{f}}}^{WSE}$, typically is not sufficiently large for proper application of the Central Limit Theorem. In this case, the upper and lower bound of $N_{\Omega_2^{\hat{f}}}^{WSE}$ can be estimated through Monte Carlo simulations. Denote the probability distribution of $N_{\Omega_2^{\hat{f}}}^{WSE} = \sum_{i=1}^{N_{\Omega_2^{\hat{f}}}} I_i^{\hat{f}}$ as $\boldsymbol{\Theta}$, which is the summation of a number of Bernoulli distributions. The upper and lower bound of $N_{\Omega_2^{\hat{f}}}^{WSE}$ can be estimated as:

$$N_{\Omega_2^{\hat{f}}}^{WSE} \in \left[\boldsymbol{\Gamma}^{-1}\left(\frac{\alpha}{2}\right), \quad \boldsymbol{\Gamma}^{-1}\left(1 - \frac{\alpha}{2}\right)\right] \tag{32}$$

where $\boldsymbol{\Gamma}^{-1}$ is the inverse of the cumulative density function(CDF) of $\boldsymbol{\Theta}$. For confidence level $\alpha$, the confidence interval of $N_{\Omega_2^{\hat{f}}}^{WSE}$ and $N_{\Omega_2^{\hat{s}}}^{WSE}$ can be calculated as:

$$N_{\Omega_2^{\hat{f}}}^{WSE} \in \left[\boldsymbol{\Gamma}^{-1}\left(\frac{\alpha}{2}\right), \quad \boldsymbol{\Gamma}^{-1}\left(1 - \frac{\alpha}{2}\right)\right], \qquad N_{\Omega_2^{\hat{s}}}^{WSE} \in \left[\mu_{\Omega_2^{\hat{s}}} - \alpha_{ci}\sigma_{\Omega_2^{\hat{s}}}, \quad \mu_{\Omega_2^{\hat{s}}} + \alpha_{ci}\sigma_{\Omega_2^{\hat{s}}}\right] \tag{33}$$

where $\alpha_{ci} = 1.96$ for 95% confidence level. It should be noted that points with wrong sign estimation in $\Omega_2^{\hat{s}}$ increase $N_{\Omega_2^{\hat{f}}}$, while points with wrong sign estimation in $\Omega_2^{\hat{f}}$ decrease $N_{\Omega_2^{\hat{f}}}$. Therefore, the true number of fail design points in $\Omega_2$, $N_{\Omega_2^f}$ can be estimated as:

$$N_{\Omega_2^f} = N_{\Omega_2^{\hat{f}}} + N_{\Omega_2^{\hat{s}}}^{WSE} - N_{\Omega_2^{\hat{f}}}^{WSE} \tag{34}$$

where $N_{\Omega_2^{\hat{f}}}$ is known, and the upper and lower bounds of $N_{\Omega_2^{\hat{s}}}^{WSE}$ and $N_{\Omega_2^{\hat{f}}}^{WSE}$ are available according to the Eq. (33). Considering two extreme cases where $N_{\Omega_2^{\hat{s}}}^{WSE}$ reaches its upper bound and $N_{\Omega_2^{\hat{f}}}^{WSE} = 0$, and $N_{\Omega_2^{\hat{s}}}^{WSE} = 0$ and $N_{\Omega_2^{\hat{f}}}^{WSE}$ reaches its upper bound, a conservative estimate for the range of $N_{\Omega_2^f}$ represented by $N_{\Omega_2^f}^R$ is as follows:

$$N_{\Omega_2^f}^R \in \left[N_{\Omega_2^{\hat{f}}} - \boldsymbol{\Gamma}^{-1}\left(1 - \frac{\alpha}{2}\right), \quad N_{\Omega_2^{\hat{f}}} + \left(\mu_{\Omega_2^{\hat{s}}} + \alpha_{ci}\sigma_{\Omega_2^{\hat{s}}}\right)\right] \tag{35}$$

Subsequently, based on Eq. (26), the maximum error rate, $\hat{\epsilon}_{max}$, can be conservatively estimated as:

$$\epsilon \cong \left|\frac{N_{\Omega_2^{\hat{f}}} - N_{\Omega_2^f}}{N_{\Omega_1^{\hat{f}}} + N_{\Omega_2^f}}\right| \leq \max_{I_{\Omega_2} \in N_{\Omega_2^f}^R} \left(\left|\frac{N_{\Omega_2^{\hat{f}}} - I_{\Omega_2}}{N_{\Omega_1^{\hat{f}}} + I_{\Omega_2}}\right|\right) = \hat{\epsilon}_{max} \tag{36}$$

The estimation of the maximum error rate is integrated in REAK as explained in detail next. As noted before, events of wrong sign estimation for candidate design points by Kriging model are correlated.

However, considering the high computational demands associated with the incorporation of correlated models, this study does not consider correlations in wrong sign estimations.

### 3.3. REAK Algorithm

As stated before, REAK involves three major processes including novel effective sampling regions and maximum error rate control that are integrated into AK-MCS. REAK uses AK-MCS to perform reliability analysis, employs ESR to adaptively update the sampling regions and treats estimated maximum error rate as stopping criterion within the effective sampling regions. A flowchart that illustrates REAK algorithm is shown in Fig. 3. The primary steps of the algorithm are elaborated below:

- **Step 1**: *Generating initial candidate design samples.* In this step, $N_{MCS}$ candidate design samples are generated by Latin Hypercube Sampling (*LHS*); the set of these samples is denoted as *S*.

- **Step 2**: *Initial setting of $\alpha$ and effective sampling regions.* A reasonable initial value for $\alpha$ denoted as $\alpha_{initial}$ can be determined using the following quadratic equation:

$$\alpha_{intial} = \gamma \cdot \epsilon_{thr} \cdot n_r^2 \tag{37}$$

where $\epsilon_{thr}$ is the threshold for the error rate, $n_r$ is the number of random variables and $\gamma$ is a constant coefficient. Note that $\gamma$ is introduced to define only the initial value of $\alpha$ and not the final value of $\alpha$. For example, if the target $\alpha$ is 0.2, starting with $\alpha = 1$ and $\alpha = 1.44$ corresponding to $\gamma = 5$ and $\gamma = 6$, respectively, will both lead to the target $\alpha = 0.2$. Therefore, $\gamma$ is not fixed because the change in $\alpha$ in fact reflects the change in $\gamma$, which is a main difference between REAK and ISKRA. It should be noted that $\gamma$ should not very large. The reason is that if $\gamma$ is very large, the effective sampling regions (ESR) will be too small to capture important information regarding the performance function. This subsequently results in unnecessary evaluations of the performance function. Testing experience shows that $\gamma \in (4, 6)$ can yield acceptable results. In this paper, $\gamma$ is set to 5. Table 1 shows the relation between $\alpha_{initial}$ and $\epsilon_{thr}$ for problems involving two random variable ($n_r = 2$).

**Table 1.** The relation between $\alpha_{intial}$ and $\hat{\epsilon}_{max}$ for the case of $n_r = 2$.

| $\epsilon_{thr}$ | 0.01 | 0.02 | 0.03 | 0.04 | 0.05 |
|---|---|---|---|---|---|
| $\alpha_{intial}$ | 0.2 | 0.4 | 0.6 | 0.8 | 1.0 |

- **Step 3**: *Initial training points.* Randomly select a limited number of points from *S* as initial training points for Kriging construction. Denote the training points as $\boldsymbol{x}_{tr}$.

- **Step 4**: *Kriging construction and estimation.* Construct the Kriging model with current training points $\boldsymbol{x}_{tr}$. Denote the Kriging model as $K(\boldsymbol{x})$. Construction is based on MATLAB toolbox DACE [28], with ordinary Kriging basis and Gaussian correlation function. By implementing DACE toolbox, the Kriging responses $\mu_K(\boldsymbol{x})$ and variances $\sigma_K(\boldsymbol{x})$ are obtained for every points in *S*. According to the responses $\mu_K(\boldsymbol{x})$, calculate the estimated probability of failure $\hat{P}_f$ by Eq. (1).

- **Step 5**: *Update efficient sampling regions.* In this step, effective sampling regions is defined by Eq. (17) in current *S* and is denoted as $S_{ESR}$.

- **Step 6**: *Search for the next training point.* In this stage, learning function *EFF* in Eq. (11) is used. The next training point is selected according to Eq. (18) and is denoted as $\boldsymbol{x}_{tr}^*$.

- **Step 7**: *Stopping criterion on learning process.* Determine if the stopping criterion ($max(EFF) \leq 0.001$) has been satisfied in the current iteration. Go to Step 8 if satisfied; otherwise, go back to Step 4.

- **Step 8**: *Maximum error rate prediction.* By computing Eq. (33) ~ (36), estimate the maximum error rate $\hat{\epsilon}_{max}$.

- **Step 9**: *Stopping criterion on maximum error rate.* Check the stopping criterion ($\hat{\epsilon}_{max} \leq \epsilon_{thr}$) for maximum error rate. If the stopping criterion is not satisfied, then expand the effective sampling region by updating ESR as follows:

$$\alpha^{new} = \alpha - \Delta_\alpha \tag{38}$$

$$P_{\{\rho(x)>\rho_{thr}\}} = \alpha^{new} \hat{P}_f \tag{39}$$

where $\Delta_\alpha$ determines additional number of points to be added to ESR. Different values for $\Delta_\alpha$ can be selected; however, large values of this parameter may lead to accuracies higher than required and therefore to unnecessary computational demand. For instance, if $\Delta\alpha$ is set to be large, $\hat{\epsilon}_{max}$ can directly jump from 0.1 to 0.01 after a large number of points are accepted in ESR, while the target $\epsilon_{thr}$ is 0.05. For this purpose, $\Delta_\alpha = 0.01$ is adopted in this paper. Other definitions of expanded ESR remains the same as in Eq. (17). After ESR is updated, go back to Step 5. If the maximum error rate stopping criterion is satisfied, go to Step 10.

- **Step 10**: *Estimating coefficient of variation of failure probability.* In this step, the following equation is used to check whether the population of *S* is sufficient for failure probability estimation:

$$COV_{P_f} = \sqrt{\frac{1-\hat{P}_f}{\hat{P}_f N_{MCS}}} \leq COV_{thr} \tag{40}$$

where $COV_{thr}$ is the threshold of variation of $\hat{P}_f$. An acceptable value for this threshold is 0.05. If Eq. (40) is satisfied, then go to Step 11. If not, it means that the number of $N_{MCS}$ is not sufficient and additional number of candidate design samples should be added to *S* and Step 4 needs to be repeated.

- **Step 11:** *End of REAK and output $\hat{P}_f$.* Unlike the process in ISKRA [23] which only takes a fixed value for $\alpha$ e.g. 0.05, REAK algorithm adaptively decreases $\alpha$ until the constraint on the estimated maximum error rate is satisfied. As will be shown later, the target error rate ($\epsilon_{thr} = 0.05$) can be potentially achieved even for large values of $\alpha$ ($\alpha > 1$). Taking advantage of this in REAK results in considerable reduction of computational demand compared to ISKRA method. Furthermore when the target error rate is set to be very small (e.g. $\epsilon_{thr}$= 0.01 or 0.001), the reduction in the number of calls to performance function by ISKRA compared to AK-MCS becomes less significant, while the reduction by REAK is still considerable.

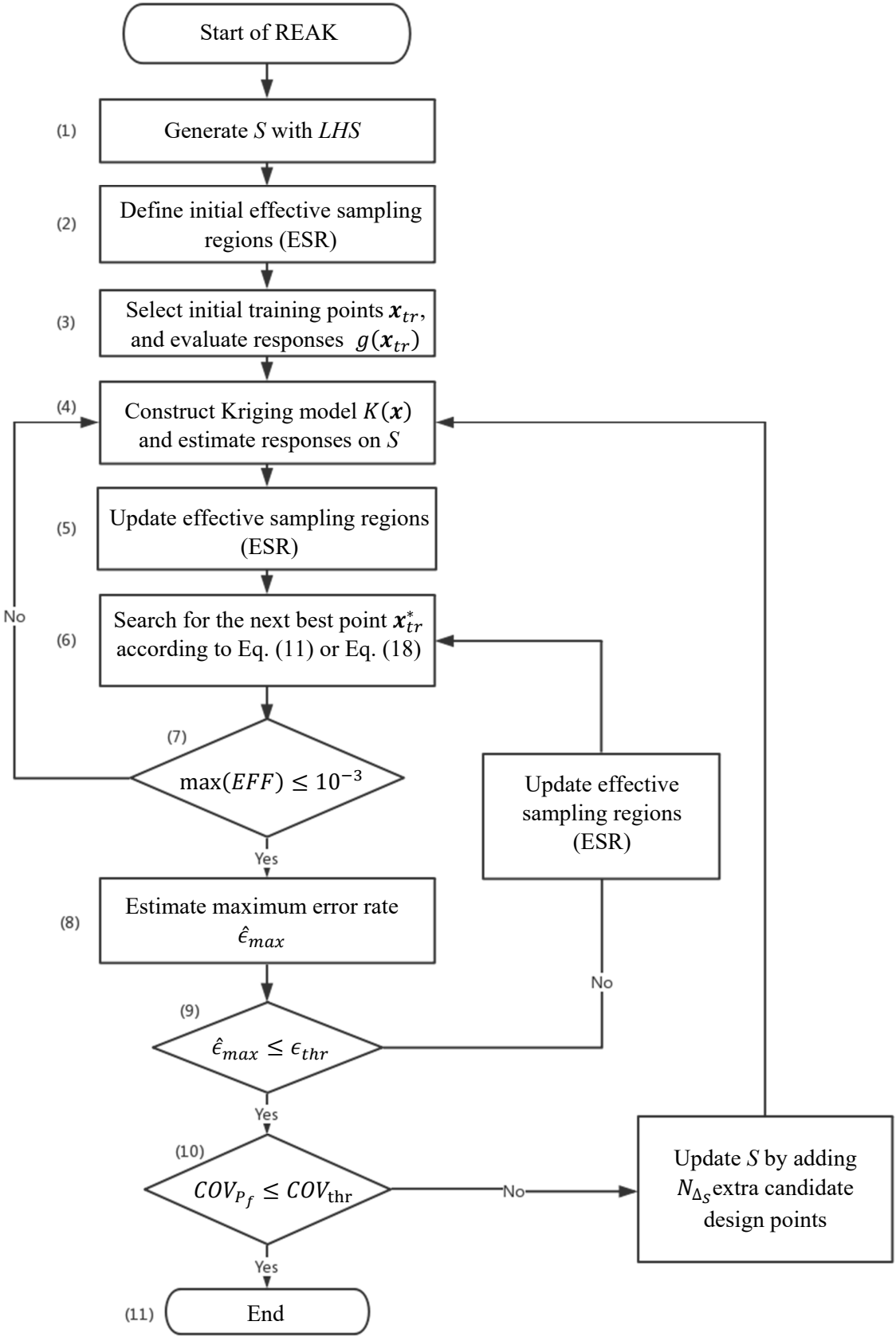

**Fig 3.** REAK flowchart

## 4. NUMERICAL INVESTIGATION

The performance of the proposed REAK algorithm is thoroughly investigated against MSC, AK-MCS, and ISKRA techniques. This evaluation is conducted for four different examples of varying complexities with respect to the extent of nonlinearity in the limit state function and the dimension as presented below.

### 4.1. Four-Boundary Series System

The first classical example is a four-boundary series system provided in [7], [9], [11], [22], [23], [33]. The contour of this problem exhibits a highly nonlinear behavior. This leads to high variations in failure probability estimation results by crude Monte-Carlo simulation. The performance function of this problem that involves two independent and identically distributed (iid) standard normal random variables $x_1$ and $x_2$ is presented as:

$$g(x_1, x_2) = \min \begin{cases} 3 + 0.1(x_1 - x_2)^2 - \dfrac{(x_1 + x_2)}{\sqrt{2}} \\ 3 + 0.1(x_1 - x_2)^2 + \dfrac{(x_1 + x_2)}{\sqrt{2}} \\ (x_1 - x_2) + \dfrac{6}{\sqrt{2}} \\ -(x_1 - x_2) + \dfrac{6}{\sqrt{2}} \end{cases} \qquad (41)$$

Results of reliability analysis using MCS, AK-MCS, ISKRA, REAK, and two other variants of REAK including REAK* and REAK** are presented in Table 2. These variants are designed to examine the contributions of two different components in REAK: ESR and stopping criterion. In REAK*, $\alpha = 0.2$ is set to be fixed and the proposed error rate-based stopping criterion in REAK is applied. First, the initial ESR is defined and kept the same throughout the process. Since the ESR remains the same, the estimated maximum error rate will not improve considerably when the next best training point is selected in ESR. Thus, the next best training point is searched globally. The above process is repeated until the stopping criterion is satisfied. On the contrary, in REAK**, $\alpha = 1$ is incrementally reduced until $\alpha = 0.2$. No error rate-based stopping criterion is applied in this variant. In this example, the threshold for coefficient of variation of failure probability estimate using MCS is set to be 1.5%. The comparisons are all based on the same candidate design samples $S$ and initial set of training points. Results of each of these methods include the number of calls to performance function, $N_{call}$, estimated probability of failure, coefficient of variation of estimated probability of failure, $COV_{P_f}$, estimated maximum error rate, $\hat{\epsilon}_{max}$, and the true error rate, $\epsilon$. Results indicate that the proposed method REAK significantly reduces the number of calls to the performance function when compared with ISKRA and AK-MCS methods. In ISKRA algorithm, the estimated maximum error rate $\hat{\epsilon}_{max}$ is considered as 0.05 when $\alpha$ is 0.05 [23]. However in REAK, for the same target error rate of 0.05, $\alpha$ can be significantly larger due to the implemented adaptive process. As a result, REAK is able to identify significantly larger number of candidate design samples that are not important for failure probability estimation and therefore excluded them from the estimation process as seen in Fig. 4 (a) and (b); this consequently reduces the number of calls to the performance function as seen in the results. Additionally, the estimated Kriging models in Fig. 4 (a) and (b) show that the limit state function estimated by Kriging (shown by connected red points) does not need to be refined perfectly to match the true limit state function (shown by connected black points); in fact, the extent of required refinement depends on the contribution of the regions to the failure probability and the maximum error rate threshold. This feature can help researchers reduce the number of calls to computationally demanding performance models when the goal is to achieve a target error rate.

Compared with AK-MCS, REAK is considerably more efficient. The property of efficiency can be seen here in terms of the number of calls to the performance function. When the threshold of error rate $\epsilon_{thr}$ is set to 0.05, the number of required calls by REAK and AK-MCS are 60 and 125, respectively, while both methods achieve high levels of accuracy (2.16% for REAK and 0.01% for AK-MCS) and satisfy the threshold limit for error. Comparing the performance of the two variants of REAK i.e. REAK* and REAK**, it appears that both the proposed error rate-based stopping criterion and ESR contribute to the reduction in the number of calls to the performance function. However, the contribution by the stopping criterion appears to be more significant. Moreover, REAK provides a reliable upper bound estimate of the maximum error rate; a feature that is missing in AK-MCS. As the error threshold decreases, the number of performance function evaluations increases leading to a tighter upper bound since the accuracy of the error rate estimate that is derived based on the Central Limit Theorem increases. Concerning Eq. (16), one may argue that the additional 65 (=125-60) function evaluations required to improve the accuracy from $\epsilon = 0.0216$ to $\epsilon = 0.0007$ is not justifiable. This is a common practice in engineering applications where the acceptable maximum error in estimated reliabilities is not very small, and the 5% limit is often considered to be appropriate. Consequently, REAK is a very powerful and efficient technique as its error prediction feature allows users to set error thresholds appropriate to the particular application.

**Table** 2. Reliability analysis results for MCS, AK-MCS, ISKRA, REAK, REAK* and REAK** for different error rate thresholds. The threshold $COV_{\text{thr}}$ for $COV_{P_f}$ is 0.015, the initial number of candidate design points for AK-MCS and REAK is $N_S = 10^5$ with $N_{\Delta_S} = 10^5$ and $\gamma$ in Eq. 37 is 20 in this case.

| $\epsilon_{thr}$ | Methodology | $N_{call}$ | $\hat{P}_f(COV_{P_f})$ | $\hat{\epsilon}_{max}$ | $\epsilon$ |
|---|---|---|---|---|---|
| | Monte Carlo | $10^6$ | $4.498 \times 10^{-3}$ (1.5%) | - | - |
| | AK-MCS+EFF | $12 + 113$ | $4.495 \times 10^{-3}$ | No Estimate | 0.0007 |
| 0.05 | REAK | $12 + 48$ | $4.401 \times 10^{-3}$ | 0.0375 | 0.0216 |
| | ISKRA | $12 + 108$ | $4.498 \times 10^{-3}$ | 0.05 | 0 |
| 0.03 | REAK | $12 + 54$ | $4.425 \times 10^{-3}$ | 0.0247 | 0.0162 |
| | ISKRA | $12 + 110$ | $4.498 \times 10^{-3}$ | 0.03 | 0 |
| 0.01 | REAK | $12 + 64$ | $4.478 \times 10^{-3}$ | 0.0082 | 0.0044 |
| | ISKRA | $12 + 115$ | $4.498 \times 10^{-3}$ | 0.01 | 0 |
| | REAK* | $12 + 68$ | $4.473 \times 10^{-3}$ | 0.0062 | 0.0057 |
| | REAK** | $12 + 76$ | $4.511 \times 10^{-3}$ | No Estimate | 0.0029 |

As stated before, in order to have a fair comparison among various considered reliability analysis methods, the set of design samples, $S$, is kept the same for all methods. However, this set is generated randomly and the performance of sampling-based methods such as those considered in this study may be different from one set to another set. To examine such effects, the expected performance of reliability algorithms is investigated here. Due to large number of evaluations involved, the threshold for $COV_{P_f}$ is relaxed and chosen as 0.05. The initial number of candidate design points for AK-MCS and REAK is $N_S = 10^4$ with $N_{\Delta_S} = 10^4$ as the increment. The average number of calls to performance function is denoted as $\bar{N}_{call}$, the average estimated maximum error rate is $\bar{\epsilon}_{max}$ and the average true error rate is $\bar{\epsilon}$. The number of simulations to determine the average performance is 50. $\bar{N}_{call}$ vs $\alpha(\epsilon_{thr})$ in ISKRA, $\bar{\epsilon}$ vs $\alpha(\epsilon_{thr})$ in ISKRA, $\bar{N}_{call}$ vs $\epsilon_{thr}$ in REAK and $\bar{\epsilon}$ & $\bar{\epsilon}_{max}$ vs $\epsilon_{thr}$ in REAK are shown in in Fig. 5 (a)-(d), respectively. Additionally, the mean and coefficient of variation (C.O.V) of $N_{call}$ and $\hat{\epsilon}_{max} - \epsilon$ for ISKRA and REAK for three different error rate thresholds $\epsilon_{thr} = 0.05$, $\epsilon_{thr} = 0.03$ and $\epsilon_{thr} = 0.01$ are presented in Table 3.

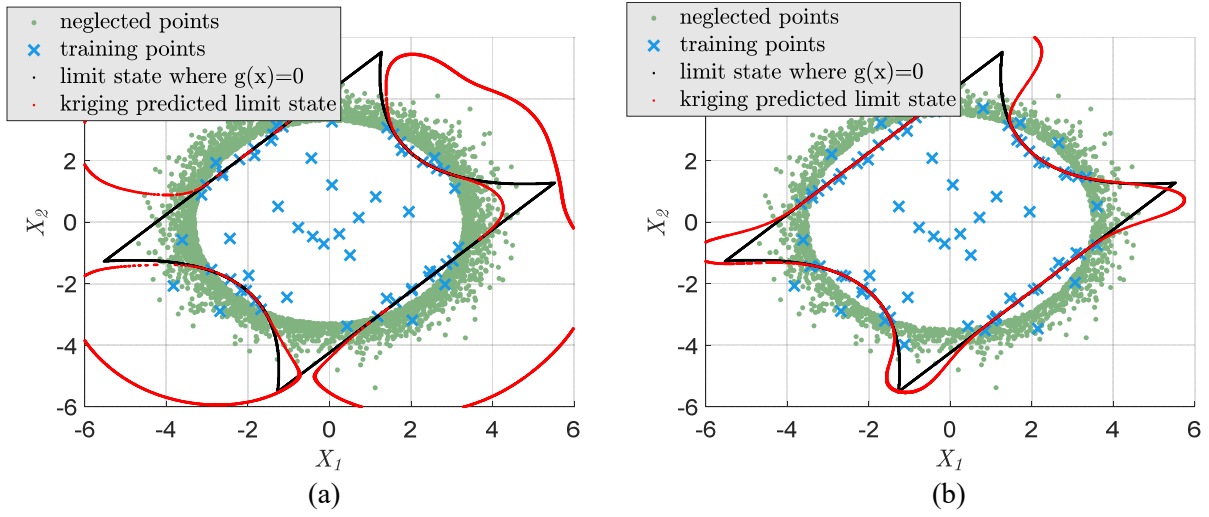

**Fig**. 4. (a) Limit state of Kriging model for $\epsilon_{thr}$= 0.05 and (b) Limit state of Kriging model for $\epsilon_{thr}$= 0.01

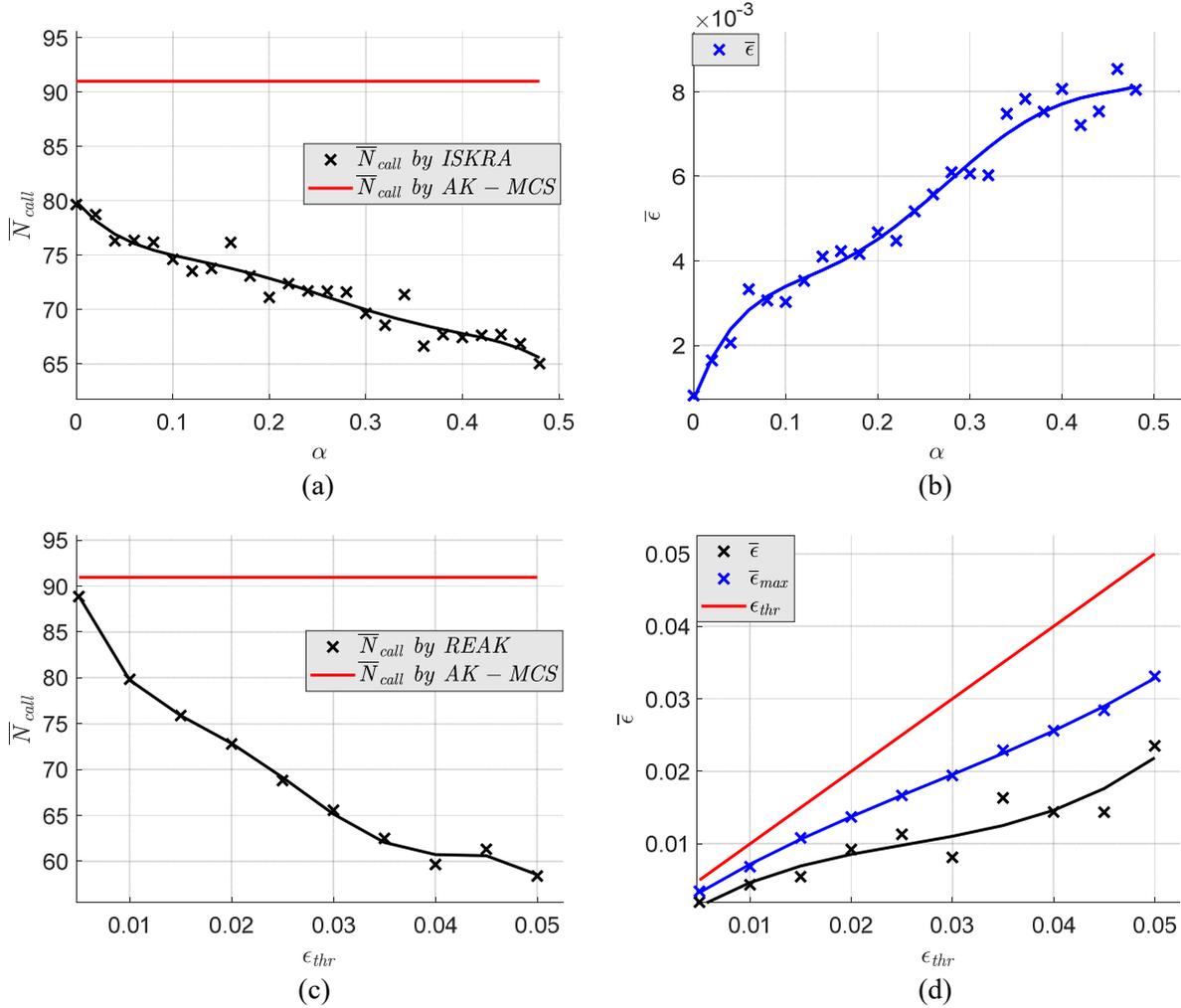

**Fig**. 5. (a) $\overline{N}_{call}$ vs $\alpha = \epsilon_{thr}$ in ISKRA, (b) $\bar{\epsilon}$ vs $\alpha = \epsilon_{thr}$ in ISKRA, (c) $\overline{N}_{call}$ vs $\epsilon_{thr}$ in REAK, and (d) $\bar{\epsilon}$ and $\bar{\epsilon}_{max}$ vs $\epsilon_{thr}$ in REAK

From the presented results, it can be observed that:
- The average number of calls to performance function for REAK, ISKRA and AK-MCS methods are 58.36, 74.60 and 90.96, respectively, for the case when the error rate threshold $\epsilon_{thr}$ is 0.05. This shows that REAK reduces $\bar{N}_{call}$ considerably compared to ISKRA and AK-MCS. Even for the higher requirement of accuracy of $\epsilon_{thr}$= 0.01, the average number of calls to performance function for REAK is 79.84 which is lower than 90.96 for AK-MCS. Hence, REAK is considerably more efficient than ISKRA and AK-MCS.

- Generally, as the threshold for error rate $\epsilon_{thr}$ increases, the average number of calls to performance function $\bar{N}_{call}$ decreases. This point is obvious in Fig. 5 (a) and (c). It should be noted that the constant coefficient $\alpha$ is equal to $\epsilon_{thr}$ in ISKRA approach.

- The average true error rate $\bar{\epsilon}$ increases as $\epsilon_{thr}$ increases. In ISKRA algorithm, $\bar{\epsilon}$ is significantly lower than $\alpha$ or $\epsilon_{thr}$. For example when $\alpha$ is 0.05, $\bar{\epsilon}$ is 0.002 as seen in Fig. 5 (b). Therefore, ISKRA cannot accurately estimate the true error rate. However in REAK algorithm, the average estimated maximum error rate $\bar{\epsilon}_{max}$ is close to but larger than $\bar{\epsilon}$ and yet smaller than the threshold $\epsilon_{thr}$; in other words $\bar{\epsilon} \leq \bar{\epsilon}_{max} \leq \epsilon_{thr}$, as seen in Fig. 5 (d). This indicates that REAK's estimate of maximum error is reliable.

- Results for the mean of $\hat{\epsilon}_{max} - \epsilon$ in Table 3 indicate that on average the estimated maximum error rates $\hat{\epsilon}_{max}$ are greater than the true error rates $\epsilon$. In addition, for the case of $\epsilon_{thr} = 0.01$, the C.O.V of $\hat{\epsilon}_{max} - \epsilon$ is 0.547, which means that the probability $\epsilon \leq \hat{\epsilon}_{max}$ is $1 - 2\Phi\left(-\frac{1}{0.547}\right) = 1 - 2\Phi(-1.74) = 91.80\%$ assuming that $\hat{\epsilon}_{max} - \epsilon$ follows a normal distribution. For $\epsilon_{thr} = 0.03$ and $\epsilon_{thr} = 0.01$, the probability of $\epsilon \leq \hat{\epsilon}_{max}$ is 97.27% and 93.25%, respectively. These results are close to the set confidence level of 95%.

**Table 3.** The performance of REAK, ISKRA and AK-MCS with 50 simulations for example 1.

| $\epsilon_{thr}$ | Methodology | Mean of $N_{call}$ | C.O.V of $N_{call}$ | Mean of $\hat{\epsilon}_{max} - \epsilon$ | C.O.V of $\hat{\epsilon}_{max} - \epsilon$ |
|---|---|---|---|---|---|
| | Monte Carlo | $1 \times 10^5$ | - | - | - |
| | AK-MCS+EFF | 90.96 | 0.137 | - | - |
| 0.05 | REAK | 58.36 | 0.212 | 0.016 | 0.575 |
| | ISKRA | 74.60 | 0.072 | 0.047 | 0.057 |
| 0.03 | REAK | 65.54 | 0.168 | 0.013 | 0.453 |
| | ISKRA | 76.34 | 0.086 | 0.027 | 0.087 |
| 0.01 | REAK | 79.84 | 0.140 | 0.003 | 0.547 |
| | ISKRA | 79.64 | 0.148 | 0.007 | 0.221 |

## 4.2. Modified Rastrigin Function

The second example for limit state function is modified Rastrigin function. It contains non-convex features, which require a significant number of calls to performance function [2], [10], [21] in order to arrive at acceptable surrogate models [7], [9], [22]. The performance function is as follows:

$$g(x_1, x_2) = 10 - \sum_{i=1}^{2}\left(x_i^2 - 5\cos(2\pi x_i)\right) \quad (42)$$

where $x_i$s are mutually independent standard normal random variables. Similar to the first example, MCS, AK-MCS, ISKRA and REAK are implemented in this case to compare their performance. Results for individual simulation and average performance are given in in Table 4 and Table 5, respectively.

Furthermore, the true limit state function $g(x)$ and its Kriging estimate $\hat{g}(x)$ are shown in Fig. 6 (a) and (b) for different $\epsilon_{thr}$ values. It is shown that the Kriging estimation of the limit state based on REAK is well refined in the effective sampling regions (ESR) but not outside of ESR. As seen in Table 4, when $\epsilon_{thr}$ is 0.05, REAK is able to evaluate $\hat{P}_f$ with 230 training points, while AK-MCS and ISKRA require 520 and 559 points, respectively. When $\epsilon_{thr}$ is 0.03, $N_{call}$ for REAK is 276 which is considerably less than $N_{call}$ of 574 for ISKRA and 520 for AK-MCS. Moreover, REAK's estimate of the true error $\hat{\epsilon}_{max}$=0.51% is very close to the true error $\epsilon$=0.43% for the case of $\epsilon_{thr}$= 0.01.

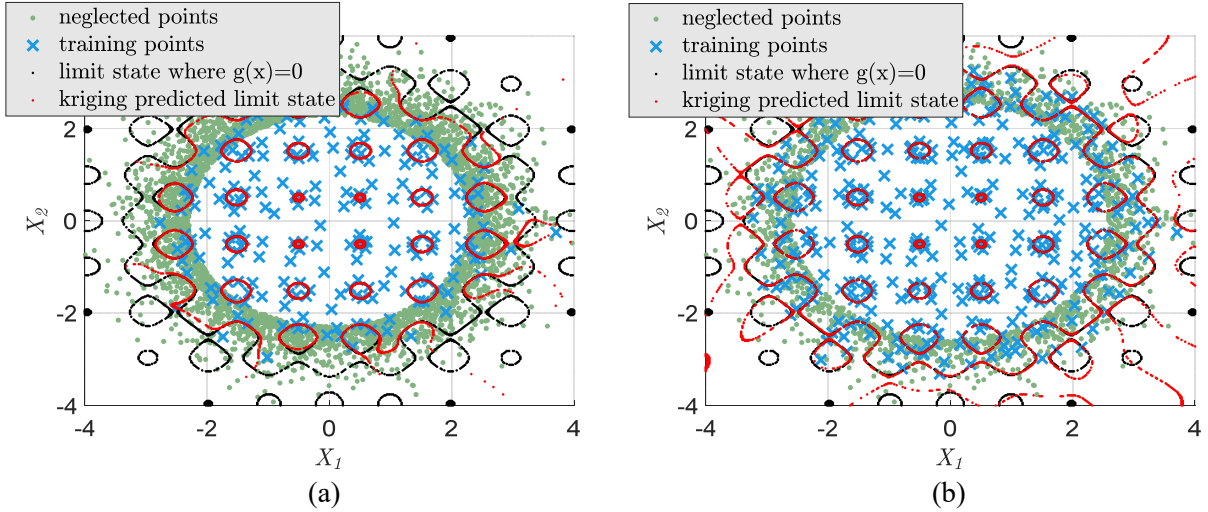

**Fig**. 6 (a) Limit state of Kriging model for $\epsilon_{thr}$= 0.05, and (b) Limit state of Kriging model for $\epsilon_{thr}$= 0.01.

**Table 4.** Reliability analysis results using MCS, AK-MCS, ISKRA and REAK methods for different error rate thresholds. The threshold $COV_{thr}$ for $COV_{P_f}$ is 0.015, the initial number of candidate design points for AK-MCS and REAK is $N_S = 10^4$ with $N_{\Delta_S} = 10^4$. Other parameters are the same as those in Table 2.

| $\epsilon_{thr}$ | Methodology | $N_{call}$ | $\hat{P}_f (COV_{P_f})$ | $\hat{\epsilon}_{max}$ | $\epsilon$ |
|---|---|---|---|---|---|
| | Monte Carlo | $6 \times 10^4$ | $7.308 \times 10^{-2}$ (1.5%) | - | - |
| | AK-MCS+EFF | 12 + 508 | $7.308 \times 10^{-2}$ | No Estimate | 0 |
| 0.05 | REAK | 12 + 218 | $7.116 \times 10^{-2}$ | 0.0438 | 0.0263 |
| | ISKRA | 12 + 547 | $7.297 \times 10^{-2}$ | 0.05 | 0.0015 |
| 0.03 | REAK | 12 + 264 | $7.157 \times 10^{-2}$ | 0.0234 | 0.0207 |
| | ISKRA | 12 + 562 | $7.308 \times 10^{-2}$ | 0.03 | 0 |
| 0.01 | REAK | 12 + 389 | $7.277 \times 10^{-2}$ | 0.0051 | 0.0043 |
| | ISKRA | 12 + 581 | $7.308 \times 10^{-2}$ | 0.01 | 0 |

The results for the average performance of methods based on 50 simulations are presented in Table 5. For all presented cases, it is observed that the probability of $\hat{\epsilon}_{max} \geq \epsilon$ with success estimate is close to 95%. For example, the probability of successful estimation of $\hat{\epsilon}_{max} \geq \epsilon$ is 94.55% when $\epsilon_{thr}$ is 0.03. Generally, through integration of adaptive expansion of effective sampling regions and error rate control, REAK successfully reduces the number of calls to performance function and offers a reliable estimate of the maximum error rate.

**Table 5.** The performance of REAK, ISKRA and AK-MCS with 50 simulations for example 2.

| $\epsilon_{thr}$ | Methodology | Mean of $N_{call}$ | C.O.V of $N_{call}$ | Mean of $\hat{\epsilon}_{max} - \epsilon$ | C.O.V of $\hat{\epsilon}_{max} - \epsilon$ |
|---|---|---|---|---|---|
| | Monte Carlo | $6 \times 10^4$ | - | - | - |
| | AK-MCS+EFF | 510.40 | 0.126 | - | - |
| 0.05 | REAK | 229.84 | 0.062 | 0.023 | 0.443 |
| | ISKRA | 508.64 | 0.082 | 0.047 | 0.063 |
| 0.03 | REAK | 275.32 | 0.075 | 0.012 | 0.520 |
| | ISKRA | 512.2 | 0.094 | 0.026 | 0.057 |
| 0.01 | REAK | 393.46 | 0.064 | 0.002 | 0.572 |
| | ISKRA | 517.44 | 0.090 | 0.008 | 0.147 |

### 4.3. Non-linear Oscillator

To demonstrate the performance of REAK for high dimensional problems, an undamped nonlinear single degree of freedom system with six random (as shown in Fig. 7) variables is investigated in this section. Details of this model can be seen in [7], [9], [33]–[35]. The performance function is describe below:

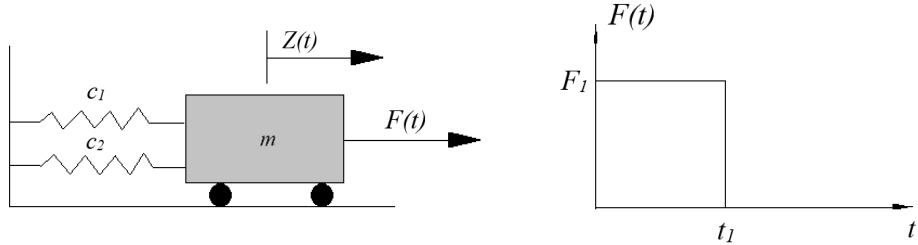

**Fig**. 7 Example 3, non-linear oscillator

$$g(c_1, c_2, m, r, t_1, F_1) = 3r - \left| \frac{2F_1}{m\omega_0^2} \sin\left(\frac{\omega_0 t_1}{2}\right) \right| \tag{43}$$

where $\omega_0 = \sqrt{\frac{c_1+c_2}{m}}$ is the frequency of the system. The description of six random variables is listed in Table 6, the results of individual simulations are summarized in Table 7 and results with 50 simulations are presented in Table 8.

**Table 6.** Random variables used in example 3.

| Random variable | Distribution type | Mean | Standard Deviation |
|---|---|---|---|
| $m$ | Normal | 1 | 0.05 |
| $c_1$ | Normal | 1 | 0.1 |
| $c_2$ | Normal | 0.1 | 0.01 |
| $r$ | Normal | 0.5 | 0.05 |
| $F_1$ | Normal | 1 | 0.2 |
| $t_1$ | Normal | 1 | 0.2 |

Results in Table 7 indicate that REAK significantly reduces the number of calls to performance function for all cases of error thresholds. Moreover, it accurately predicts the true error rate; a feature that is not available in ISKRA and AK-MCS. As also shown in Table 8, the probability of successful estimate of $\hat{\epsilon}_{max} \geq \epsilon$ is reasonable according to the confidence level in Eq. (33). Results of this example therefore represent the capability of REAK in handling reliability problems involving multiple random variables.

**Table 7.** Reliability analysis results for example 3 using MCS, AK-MCS, ISKRA and REAK methods for different error rate thresholds. The threshold $COV_{\text{thr}}$ for $COV_{P_f}$ is 0.022, the initial number of candidate design points for AK-MCS and REAK is $N_S = 10^4$ with $N_{\Delta_S} = 10^4$. Other parameters are the same as those in Table 2.

| $\epsilon_{thr}$ | Methodology | $N_{call}$ | $\hat{P}_f(COV_{P_f})$ | $\hat{\epsilon}_{max}$ | $\epsilon$ |
|---|---|---|---|---|---|
| | Monte Carlo | $7 \times 10^4$ | $2.847 \times 10^{-2}$ (2.2%) | - | - |
| | AK-MCS+EFF | $12 + 49$ | $2.844 \times 10^{-2}$ | No Estimate | 0.0010 |
| 0.05 | REAK | $12 + 18$ | $2.860 \times 10^{-2}$ | 0.0162 | 0.0045 |
| | ISKRA | $12 + 48$ | $2.849 \times 10^{-2}$ | 0.05 | 0.0005 |
| 0.03 | REAK | $12 + 22$ | $2.864 \times 10^{-2}$ | 0.0112 | 0.0060 |
| | ISKRA | $12 + 48$ | $2.849 \times 10^{-2}$ | 0.03 | 0.0005 |
| 0.01 | REAK | $12 + 28$ | $2.841 \times 10^{-2}$ | 0.0040 | 0.0020 |
| | ISKRA | $12 + 46$ | $2.846 \times 10^{-2}$ | 0.01 | 0.0005 |

**Table 8.** The performance of REAK, ISKRA and AK-MCS with 50 simulations for example 3.

| $\epsilon_{thr}$ | Methodology | Mean of $N_{call}$ | C.O.V of $N_{call}$ | Mean of $\hat{\epsilon}_{max} - \epsilon$ | C.O.V of $\hat{\epsilon}_{max} - \epsilon$ |
|---|---|---|---|---|---|
| | Monte Carlo | $7 \times 10^4$ | - | - | - |
| | AK-MCS+EFF | 60.56 | 0.092 | - | - |
| 0.05 | REAK | 28.06 | 0.020 | 0.013 | 0.481 |
| | ISKRA | 53.70 | 0.081 | 0.048 | 0.034 |
| 0.03 | REAK | 30.06 | 0.062 | 0.007 | 0.522 |
| | ISKRA | 55.82 | 0.059 | 0.028 | 0.047 |
| 0.01 | REAK | 35.24 | 0.123 | 0.002 | 0.498 |
| | ISKRA | 56.84 | 0.082 | 0.007 | 0.193 |

### 4.4. Modified Cantilever Tube

The forth example is a modified engineering problem with nine random variables. The original example is presented in [23], [36]. To include a diverse set of random variables, three types of distributions are considered: Normal, Gumbel and Uniform distributions. The cantilever tube is illustrated in Fig. 8. Three external forces $F_1$, $F_2$ and $P$ and one torsion $T$ is applied to the cantilever structure. The performance function is defined as below:

$$g(\mathrm{x}) = \sigma_{cap} - \sigma_{max} \tag{44}$$

Where $\sigma_{cap}$ is the strength capacity and $\sigma_{max}$ is the maximum von Mises stress calculated as:

$$\sigma_{max} = \sqrt{\sigma_x^2 + 3\tau_{zx}^2} \tag{45}$$

$$\sigma_x = \frac{P + F_1 sin\theta_1 + F_2 sin\theta_2}{A} + \frac{Mc}{I} \tag{46}$$

where $\theta_1 = 5°$, $\theta_2 = 10°$, and $A$ is the tube area. Moreover, $\tau_{zx}$ is the torsional stress, M is the bending moment, $c$ is the radius and $I$ is the moment inertia. These variables are calculated as:

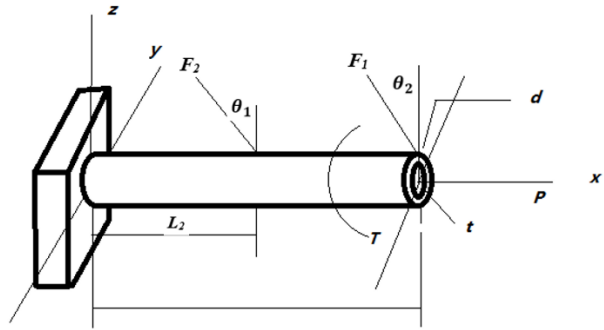

**Fig**. 8 Cantilever tube and involved variables

$$M = F_1 L_1 cos\theta_1 + F_2 L_2 cos\theta_2 \tag{47}$$

$$A = \frac{\pi}{4}[d^2 - (d-2t)^2] \tag{48}$$

$$c = \frac{d}{2} \tag{49}$$

$$I = \frac{\pi}{64}[d^4 - (d-2t)^4] \tag{50}$$

$$\tau_{zx} = \frac{Td}{2J} \tag{51}$$

$$J = 2I \tag{52}$$

The distribution information of the nine involved random variables is presented in Table 9. Results of reliability analysis for this structure are summarized in Table 10 and Table 11. The candidate design samples $S$ (including added $\Delta_S$) and initial training points $x_{tr}$ are kept the same for all considered methods including MCS, REAK, ISKRA and AK-MCS for a given threshold error rate. Results show that REAK is the most efficient technique for this high dimensional problem. In fact, REAK reduces the number of calls to performance function by 33% and 21% for $\epsilon_{thr} = 0.05$, 30% and 17% for $\epsilon_{thr} = 0.03$ compared to AK-MCS and ISKRA methods, respectively. For the case of $\epsilon_{thr} = 0.01$, the mean of $N_{call}$ for REAK is larger than that for ISKRA. Based on the results of 50 simulations presented in Table. 11, the probability that $\hat{\epsilon}_{max} \geq \epsilon$ is larger than 95%. For example, the probabilities of $\hat{\epsilon}_{max} \geq \epsilon$ are 99.97%, 99.58% and 98.5% when $\epsilon_{thr}$ is 0.05, 0.03 and 0.01, respectively. These results demonstrate the capability of REAK to set target accuracies that are appropriate for engineering problems.

**Table 9.** Random variables used in example 4.

| Random variable | Distribution type | Parameter 1 | Parameter 2 |
|---|---|---|---|
| $t$ | Normal | 5 mm (μ) | 0.1 mm (σ) |
| $d$ | Normal | 42 mm (μ) | 0.5 mm (σ) |
| $F_1$ | Normal | 3.0 kN (μ) | 0.3 kN (σ) |
| $F_2$ | Normal | 3.0 kN (μ) | 0.3 kN (σ) |
| $T$ | Normal | 90.0 Nm (μ) | 9.0 Nm (σ) |
| $\sigma_{cap}$ | Normal | 220.0 MPa (μ) | 22.0 MPa (σ) |
| $L_1$ | Uniform | 119.75 mm (lb) | 120.25 mm (ub) |
| $L_2$ | Uniform | 59.75 mm (lb) | 60.25 mm (ub) |
| $P$ | Gumbel | 27.0kN (μ) | 2.7kN (σ) |

*Note: μ and σ represent the mean and standard deviation, and lb and ub the lower and upper bounds of random variables, respectively.

**Table 10.** Reliability analysis results for example 4 using MCS, AK-MCS, ISKRA and REAK methods for different error rate thresholds. The threshold $COV_{thr}$ for $COV_{P_f}$ is 0.05, the initial number of candidate design points for AK-MCS and REAK is $N_S = 10^4$ with $N_{\Delta_S} = 10^4$. Other parameters are the same as those in Table 2.

| $\epsilon_{thr}$ | Methodology | $N_{call}$ | $\hat{P}_f(COV_{P_f})$ | $\hat{\epsilon}_{max}$ | $\epsilon$ |
|---|---|---|---|---|---|
| | Monte Carlo | $6 \times 10^4$ | $6.850 \times 10^{-3}$ (5%) | - | - |
| | AK-MCS+EFF | 12 + 72 | $6.850 \times 10^{-3}$ | No Estimate | 0 |
| 0.05 | REAK | 12 + 44 | $6.717 \times 10^{-3}$ | 0.0218 | 0.0195 |
| | ISKRA | 12 + 59 | $6.850 \times 10^{-3}$ | 0.05 | 0 |
| 0.03 | REAK | 12 + 47 | $6.817 \times 10^{-3}$ | 0.0145 | 0.0049 |
| | ISKRA | 12 + 59 | $6.850 \times 10^{-3}$ | 0.03 | 0 |
| 0.01 | REAK | 12 + 59 | $6.833 \times 10^{-3}$ | 0.0049 | 0.0024 |
| | ISKRA | 12 + 59 | $6.850 \times 10^{-3}$ | 0.01 | 0 |

**Table 11.** The performance of REAK, ISKRA and AK-MCS with 50 simulations for example 4.

| $\epsilon_{thr}$ | Methodology | Mean of $N_{call}$ | C.O.V of $N_{call}$ | Mean of $\hat{\epsilon}_{max} - \epsilon$ | C.O.V of $\hat{\epsilon}_{max} - \epsilon$ |
|---|---|---|---|---|---|
| | Monte Carlo | $6 \times 10^4$ | - | - | - |
| | AK-MCS+EFF | 83.12 | 0.082 | - | - |
| 0.05 | REAK | 51.72 | 0.054 | 0.021 | 0.364 |
| | ISKRA | 74.44 | 0.168 | 0.048 | 0.021 |
| 0.03 | REAK | 60.06 | 0.046 | 0.010 | 0.379 |
| | ISKRA | 74.58 | 0.044 | 0.027 | 0.037 |
| 0.01 | REAK | 80.42 | 0.057 | 0.004 | 0.458 |
| | ISKRA | 75.64 | 0.039 | 0.009 | 0.023 |

Although the proposed algorithm has shown a considerably better performance compared to existing techniques as demonstrated using several numerical examples, there are a number of areas where its performance can be further improved in the future. First, very high-dimensional reliability problems are challenging to solve using REAK. This is an inherent limitation of Kriging as model parameter estimations in this surrogate modeling approach requires multiple inversions of large covariance matrices. This challenge can be addressed through implementation of dimension reduction techniques [37]–[39]. Moreover, REAK assumes that events of wrong sign estimation by Kriging for candidate design samples are uncorrelated. For error thresholds that can be of interest in engineering applications, e.g. 5% or smaller, the estimated maximum error rate is sufficiently accurate. For higher thresholds of error, considering correlations among the responses from the Kriging model is expected to improve the performance of REAK. However this task will considerably increase the computational complexity of the approach, which must be addressed in the future. Moreover, the proposed methodology can be further enhanced through integration with techniques such as importance sampling and subset simulation to reduce the computational demand even further.

Unsupervised learning techniques can be leveraged to identify multiple training points simultaneously, which will facilitate parallelization of REAK. The extent of uncertainty, response of the performance function and location of the points are among key features that should be considered in the selection of best training points in effective sampling regions (ESR). Similar methodologies have been used in the literature [23], [40] for parallelization of reliability analysis, which can be applied to REAK.

## 5. CONCLUSION

This paper presents a highly efficient reliability analysis algorithm called Reliability analysis through Error rate-based Adaptive Kriging or REAK. This is achieved by developing an analytical approach to derive a maximum error rate for failure probability estimates using the Central Limit Theorem for non-identical random variables based on Lindeberg's condition, introducing effective sampling regions (ESR), and integrating these features for Kriging-based active learning. Two primary advantages offered by REAK over existing methods are: substantially decreasing the number of calls to the performance function and therefore reducing the computational demand, and determining and controlling the error rate of estimated probability of failure.

Four examples are investigated in this paper to demonstrate the application and capabilities of REAK compared to crude Monte Carlo Simulation (MCS), Adaptive Kriging with Monte Carlo Simulation (AK-MCS) and Improved Sequential Kriging Reliability Analysis (ISKRA) methods. From the reliability analysis results, a number of important observations are made which are summarized here: 1) the number of calls to performance function decreases as the threshold on error rate increases; 2) the derived maximum error rate is close to but larger than the true error rate; the maximum error rate converges to the true error rate as the number of training samples increases; and 3) the number of calls to performance function is always less than that required for AK-MCS and ISKRA. In fact, REAK reduced the number of calls by 52% and 50% for example 1, 56% and 59% for example 2, 51% and 50% for example 3, and 33% and 21% for example 4 compared to AK-MCS and ISKRA methods, respectively, for the error rate threshold of 5%. These three important features enable researchers to establish a balance between accuracy and computational demand considering the requirements of the problem at hand.

## ACKNOWLEDGEMENTS

This research has been partly funded by the U.S. National Science Foundation (NSF) through awards CMMI-1462183, 1563372, 1635569, and 1762918. Any opinions, findings, and conclusions or recommendations expressed in this paper are those of the authors and do not necessarily reflect the views of the National Science Foundation.